# Electronics-free, ultra-low-power, wearable sensor chip for high-frequency electromagnetic field detection

Abdul Mohizin[1†], Leon Abelmann[2†], Baeckkyoung Sung[1,3]*

1. *Biosensor Group, KIST Europe Forschungsgesellschaft mbH, 66123 Saarbrücken, Germany*
2. *Department of Microelectronics, Delft University of Technology, 2628 CD Delft, The Netherlands*
3. *Division of Energy & Environment Technology, University of Science & Technology (UST), 34113 Daejeon, Republic of Korea*

[†] These authors contributed equally to this work.
*Correspondence: sung@kist-europe.de

## Abstract

High-frequency electromagnetic fields (EMFs) are increasingly recognized either as environmental risk factors or as tools for electromagnetic attacks, which are difficult to detect in situ. Existing high-frequency EMF sensors face significant limitations related to structural simplicity, integration with mobile technology, and low energy consumption. To address these challenges, we propose a novel sensor concept based on a magnetically hybridized liquid crystal (LC) microdevice. The hybrid LC chip is designed to exhibit an optical response to external radio-frequency fields without the need for electronic components or an external power supply, relying solely on ambient light. Both sides of the chip are covered with polymer-based crossed polarizer films. The chip is filled with flexible matrices containing thermotropic LCs, such as the rod-like 4-cyano-4′-pentylbiphenyl, into which a network of thin ferromagnetic wires is embedded. The resulting field-responsive LC microdisplay operates via a simple magnetothermal mechanism, and its optical response is sufficiently strong to be visible to the naked eye.

## Introduction

Electromagnetic radiation is ubiquitous in modern environments, originating from sources such as power lines (Repacholi, 2012), communication systems (Russell, 2018) and imaging technologies (Wilmink & Grundt, 2011). In particular, high-frequency EMFs are widely utilized not only in heavy industries and military facilities, but also in mobile communication engineering, including 5G/6G wireless networks and satellite-based communications. While short-term health risks are generally minimal when field intensities remain within prescribed limits, long-term exposure to high-frequency electromagnetic radiation has been associated with detrimental effects on human health and ecosystems (D'Andrea et al., 2007; Hardell & Sage, 2008; Levitt et al., 2021).

To mitigate these risks, it is essential to ensure that EMF intensities remain within established safety guidelines (ICNIRP, 2020). Typically, industries monitor these intensities to prevent individual devices from exceeding permissible limits. However, in real-world applications, local field intensities can increase due to the superposition of multiple sources or equipment malfunction. This situation underscores the growing need for reliable and miniaturized electromagnetic radiation sensors.

Furthermore, high-frequency fields serve as the physical medium of electromagnetic pulse-based weapons, which can be used both for defense applications and, adversely, in cyberterrorism (Quisquater & David, 2005). In such use-cases, real-time detection of destructive EMFs significantly enhances the risk management capabilities in situ. Consequently, miniaturized sensors optimized for low energy consumption and wearable platforms are in high demand for command-and-control systems and battlefield operations.

Most existing high-frequency EMF sensor technologies require an external power source. In many applications, it is advantageous to focus on measuring the magnetic field component (Kramer et al., 2006). Predominant magnetic field sensing technologies rely on solid-state systems (Smirnov et al., 1991; Robbes, 2006; Cha et al., 2008; Asfour et al., 2014; Rodrigo et al., 2020), such as semiconductors, magnetic metals, and superconducting multilayers, and operate using detection electronics powered by external energy supplies. A wide range of devices based on the Hall effect, anisotropic magnetoresistance, giant magnetoresistance, and spin-dependent tunneling offer compact sensor designs with medium-to-high accuracy, making them well suited for on-site magnetic field measurements (Ripka & Janosek, 2010). In addition, vector magnetometers including fluxgate magnetometers (<10 kHz), fiber-optic magnetometers (<60 kHz), search-coil magnetometers (<1 MHz), Hall effect sensors (<1 MHz), and magnetoresistive sensors (<1 GHz) have been widely developed and employed for industrial applications (Lenz & Edelstein, 2006). Furthermore, portable sensors for on-site probing of radio-frequency electric fields have been developed using Rydberg-atom-based fiber-optic systems (Simons et al., 2018; Anderson et al., 2021).

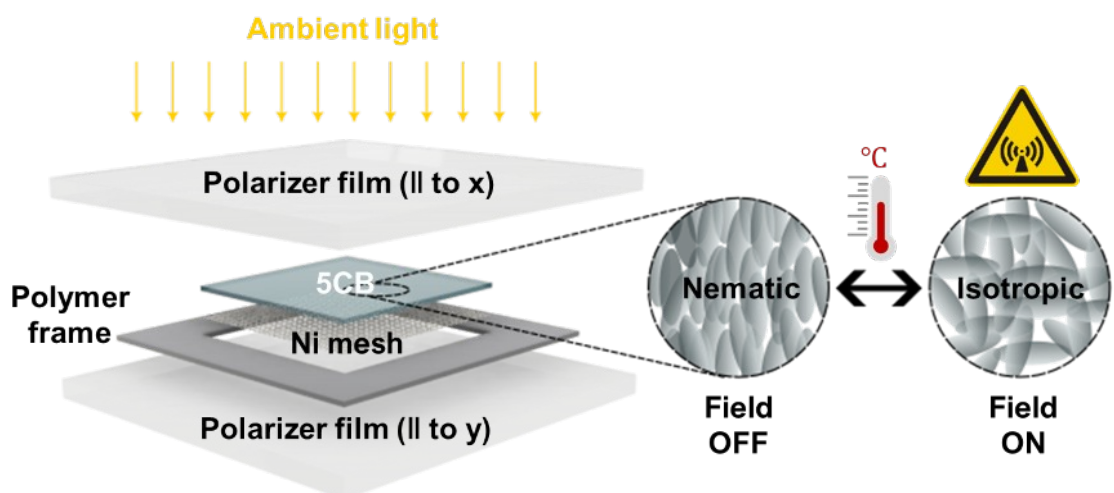

**Figure 1:** Schematic of the sensor chip for detecting high-frequency magnetic fields.

In contrast, the development of passive miniature EMF sensors—devices that derive their energy directly from the radiation source—has remained limited. Such passive, small-scale sensor systems could enable applications such as lightweight wearable devices to protect individuals or equipment at risk of overexposure, as well as low-cost indicators on devices that may emit hazardous radiation when malfunctioning. Drone warfare also needs simple onboard microsensors that are capable of detecting and alarming potential electromagnetic attacks. These sensors may exploit time-varying magnetic fields to induce electric currents in conductors, which can either power electronic devices or be converted into thermal energy.

In this study, we present a proof-of-concept sensor design, featuring an external power-free and ultra-lightweight chip format. We investigate whether the microheat generated by field-induced currents can be used to trigger a phase transition in an on-chip LC layer. This phase transition manifests as a change in LC transparency, providing a passive and energy-efficient method for detecting electromagnetic radiation. The approach converts electromagnetic energy into an optical response that can be directly observed. As the field-responsive component, a fine metal mesh is embedded within an LC-filled thin flat container based on crossed polarizers, enabling monitoring of the temperature-dependent LC phase transition in transmission. Both theoretical and experimental results demonstrate that the sensor exhibits a field sensitivity of approximately 6 mT at a radiation frequency of 125 kHz, with a response time on the order of milliseconds. These promising initial results indicate substantial potential for further optimization of this passive sensing approach.

## Materials and methods

*Sensor design*

Hybrid LC sensor chips are designed to respond passively and optically to radio-frequency EMFs without any electronics-related components or external power supply, relying solely on ambient light (either natural or artificial illumination). The chips are filled with flexible matrices comprising thermotropic LCs, such as the rod-like 4-cyano-4'-pentylbiphenyl (5CB), into which various meshes of thin ferromagnetic wires (each with diameter of 50 µm) are embedded. Both sides of the chip are laminated with thin polarizer films aligned with 90° to each other (i.e., crossed polarizers). The overall design and architecture of the sensor chip are illustrated in Fig. 1.

*Materials*

The nonchiral nematic LC compound, 4-cyano-4´-pentylbiphenyl (5CB, 98%), was purchased from Sigma-Aldrich (MO, USA). Nickel (Ni) gauze with 50-mesh and wire diameter of 50 µm (Thermo Fisher Scientific, MA, USA) was used as the baseline condition. Gauzes with 40-mesh (180-µm wire diameter), 60-mesh (130-µm wire diameter) and 100-mesh (100-µm wire diameter) were employed in parametric studies of the chip response to external fields. High-contrast linear polarizing films (XP4, Edmund Optics) were used to polarize the light transmitted through the LC cell. A 25-µL Gene Frame (Thermo Fisher Scientific) with dimensions of 10×10×0.25 mm$^3$ was used to contain the 5CB and the Ni gauze.

*Preparation of LC chips*

The polarizer films were first cut into square pieces with dimensions of 20 mm × 20 mm. Channels approximately 1 mm wide were systematically cut into a microarray slide frame (Gene Frame, Thermo Scientific) prior to attachment to one of the polarizer films. Small apertures were created in the polarizer film and aligned with the channels in the Gene Frame to enable subsequent filling of 5CB after assembly of the LC chip. The polarizer films and Ni mesh were then assembled in precise alignment with the slide frame. A total of 25 mg of 5CB was introduced into the LC cell through an opening in the assembly. Finally, the channel openings were sealed using a UV-curable resin.

*Sensor testing under applied fields*

The sensor response to alternating fields was investigated using a custom-built field generator powered by a Huttinger TIG induction heater power supply (Huttinger Elektronik, Germany). A circular coil with a diameter of 50 mm, consisting of three turns and a 50-mm pitch, was used to generate the magnetic field under baseline conditions. The field magnitude was controlled by adjusting the supply voltage, while the frequency was held constant (at 125 kHz) for a given induction coil.

The field was applied until a complete nematic-to-isotropic phase transition occurred in the LC chip. After the transition, the field was switched off, and the LC chip was allowed to cool by natural convection at room temperature (20°C), inducing the reverse transition from the isotropic to the nematic phase. The entire process was recorded using a high-speed camera (EOS 500D, Canon, Tokyo, Japan) with the aid of an elliptical mirror (Fig. S2b). The recorded images were processed using a MATLAB (MathWorks, CA, USA) algorithm to determine the time required for the phase transitions.

A calibration study (Fig. S2c) was performed using a single-turn circular copper coil with a diameter of 40 mm. The induced voltage was measured with an oscilloscope under different input voltage conditions, and the corresponding magnetic field strength ($B_0$) was calculated using a simplified form of Faraday's law of electromagnetic induction, as expressed in Eq. 1.

$$V = -B_0 A\omega \cos(\omega t) \tag{1}$$

where $V$ is the induced voltage, $A$ is the cross-sectional area of the detection coil (m²), $t$ is time (s), and $\omega$ is the natural frequency of the system (rad/s). The calibration curve relating the applied voltage to the induced magnetic field is shown in Fig. S2d.

*Computational modeling*

A computational model was developed using the commercially available finite element software COMSOL Multiphysics 6.1 (COMSOL Inc., MA, USA) to investigate the transient response of the proposed LC chip in a spatially uniform magnetic field of intensity $B_0 \sin(\omega t)$. The computational domain used in the study is shown in Fig. S1. To reduce computational load, only a quarter of the actual geometry was simulated with symmetric boundary conditions applied. The model also employed the infinite element domain feature to approximate the surrounding air region. The effect of the slide frame was neglected, and the region was assumed to consist of the PMMA polarizing film.

The system was modeled as a classical induction problem, in which electromagnetic energy losses are converted into heat. The magnetic response of the LC chip was computed by solving Ampère's law without the displacement current in the frequency domain:

$$\nabla \times \vec{H} = \vec{j} = \sigma \vec{E} + \vec{j}_e \tag{2}$$

$$\vec{B} = \nabla \times \vec{A} \tag{3}$$

$$\vec{E} = -j\omega \vec{A} \tag{4}$$

where $\vec{H}$, $\vec{j}$, $\vec{B}$, $\vec{A}$, and $\vec{E}$ denote the magnetic field intensity (A/m), current density (A/m²), magnetic field density (T), magnetic vector potential (Tm), and electric field strength (V/m), respectively. $\sigma$ represents the electrical conductivity. The physical properties of the materials used in the computational domain are summarized in Table 1.

Table 1: Physical properties of the core materials used in the modeling

| Materials & properties | Value | Reference |
|---|---|---|
| **Nickel** | | |
| Electrical conductivity | 1.43× 10⁷ S/m | 1),2) |
| Density | 8900 kg/m³ | 2) |
| Relative permeability | 148 | Timsit, 2008 |
| Heat capacity at constant pressure | 444 J/(kg·K) | 2) |
| Thermal conductivity | 91 W/(m·K) | 2) |
| **5CB** | | |
| Electrical conductivity | Function of frequency | Ibragimov & Rzayev, 2020 |
| Relative permittivity | 11.2 | di Matteo & Ferrarini, 2002; Kittipaisalsilpa et al., 2018 |
| Density (nematic phase) | 1011 kg/m³ | Oweimreen et al., 1985 |
| Density (isotropic phase) | 1008 kg/m³ | Oweimreen et al., 1985 |
| Heat capacity at constant pressure | 1.92 J/(kg·K) | Ahlers et al., 1994 |
| Thermal conductivity (nematic phase) | 1.772 W/(m·K) | Ahlers et al., 1994 |
| Thermal conductivity (isotropic phase) | 1.5 W/(m·K) | Ahlers et al., 1994 |
| Transition temperature | 308.4 K | Ahlers et al., 1994 |
| Latent heat of enthalpy | 1.56 J/g | Van Roie et al., 2005 |
| **PMMA** | | |
| Electrical conductivity | 1.× 10⁻¹⁹ S/m | 3) |
| Density | 1190 kg/m³ | 3) |
| Relative permittivity | 3 | 3) |
| Heat capacity at constant pressure | 1420 J/(kg·K) | 3) |
| Thermal conductivity | 0.19 W/(m·K) | 3) |

1) Electrical Conductivity - Elements and other Materials, (n.d.). https://www.engineeringtoolbox.com/conductors-d_1381.html (accessed October 17, 2023)
2) Material Property Database, (n.d.). https://www.mit.edu/~6.777/matprops/nickel.htm (accessed October 17, 2023)
3) PMMA, https://www.mit.edu/~6.777/matprops/pmma.htm (accessed October 17, 2023)

The resulting heat generation and distribution were described by the following governing equations:

$$\rho C_p \frac{\partial T}{\partial t} - \nabla \cdot (k \nabla T) = Q_e \tag{5}$$

$$Q_e = \frac{1}{2} R(\vec{j} \cdot \vec{E}) \tag{6}$$

where $\rho$ represents the material density (kg/m³), $C_p$ is the specific heat capacity (J/kgK), and $k$ denotes the thermal conductivity of the medium (W/m² K). $T$ is the temperature (K) and $Q_e$ quantifies the resistive heating losses in the electromagnetic system (W/m³). An initial temperature of 20°C was assumed, and a convective boundary condition with a heat transfer coefficient of 25 W/m² K (Kosky et al., 2021) was applied.

The 5CB medium undergoes a nematic-to-isotropic (N-I) phase transition at a threshold temperature of approximately 35°C (Ahlers et al., 1994; Van Roie et al., 2005; Vardanyan et al., 2015). In this study, the N-I phase transition was approximated to occur at 35.25°C, with a latent heat of 1.56 kJ/kg (Van Roie et al., 2005). The apparent heat transfer formulation in COMSOL was used to model the transition process. This formulation solves the heat transfer equation using effective material properties for both phases to predict the phase transition. The transition was assumed to occur over a 1°C temperature interval, with $\theta_N$ representing the volume fraction of the nematic phase and $\theta_I = 1 - \theta_N$ representing the volume fraction of the isotropic phase. Thus, the effective material properties of the 5CB medium can be expressed as

$$\rho_{5CB} = \rho_N \theta_N + \rho_I \theta_I \tag{7}$$

$$k_{5CB} = k_N \theta_N + k_I \theta_I \tag{8}$$

$$C_{p,5CB} = \frac{1}{\rho_{5CB}} \left( \rho_N \theta_N C_{p,N} + \rho_I \theta_I C_{p,I} \right) + L \frac{\partial \alpha_m}{\partial T} \tag{9}$$

where the subscripts *5CB*, *N*, and *I* refer to the overall 5CB medium, the nematic phase, and the isotropic phase, respectively. *L* represents the latent heat of fusion (J/kg), and $\alpha_m$ denotes the mass fraction, which can be expressed as follows:

$$\alpha_m = \frac{1}{2} \left( \frac{\rho_I \theta_I - \rho_N \theta_N}{\rho_N \theta_N + \rho_I \theta_I} \right) \tag{10}$$

**Results**

When exposed to an external radiofrequency field with intensities exceeding 6 mT, the embedded ferromagnetic mesh within the hybrid LC matrix responds via magnetothermal effects. Specifically, magnetic induction generates localized heat at the ferromagnetic wire-mesh. As this heat diffuses into the surrounding LC matrix, the local temperature within the chip increases. Once the temperature surpasses the threshold (35°C for pure 5CB), a phase transition occurs in the host LC, from the nematic phase (quasi-1D ordered local arrangement of 5CB rods) to the isotropic phase (globally disordered state).

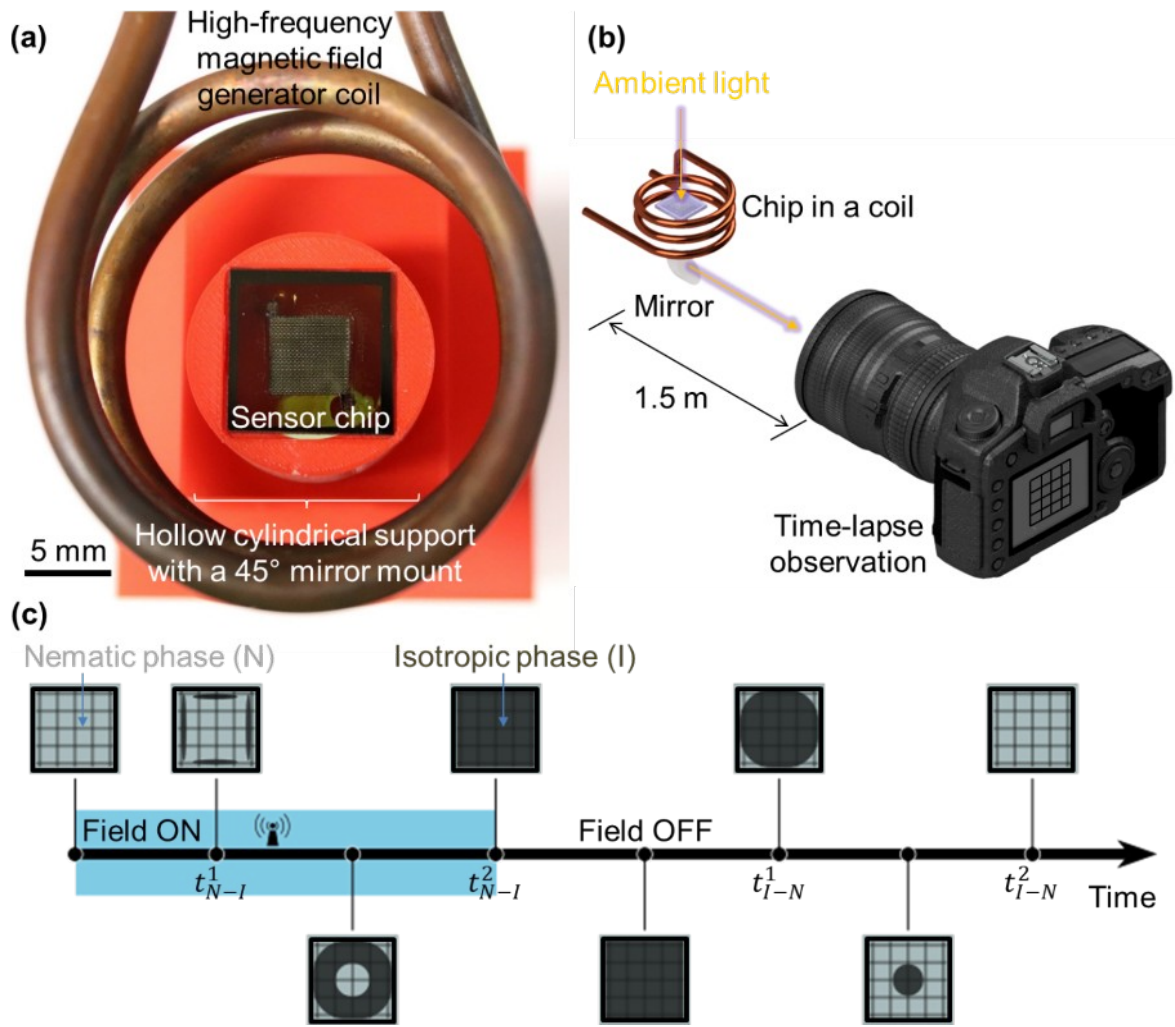

**Figure 2:** (a) Top view of the sensor chip positioned at the center of a Helmholtz coil of a high-frequency magnetic field generator. A hollow cylindrical support with an elliptical mirror mount was used for chip positioning and imaging, and a small light-emitting diode illuminator was placed behind the chip. The sensor performance was also tested under ambient daylight conditions. (b) Time-lapse imaging of the sensor's optical response was performed with a digital camera placed approximately 1.5 m from the field generator to avoid interference of the external fields with the camera electronics. (c) Schematic of the typical optical response patterns of the sensor chip, showing associated parameters under the presence and absence of external fields.

The typical response time of the hybrid matrix ranges from milliseconds to seconds. When confined within a thin, transparent chip (thickness on the order of tens of microns) and placed between crossed polarizer films (Fig. 2), the optical transmittance through the chip exhibits pronounced changes during the phase transition of the 5CB matrix. In the nematic phase (field-off condition), the system shows strong birefringence, resulting in high optical transmittance under ambient light. In contrast, under the isotropic phase (field-on condition), transmittance decreases to near zero. This process is reversible, with the system recovering within a few seconds as heat dissipates and the temperature decreases.

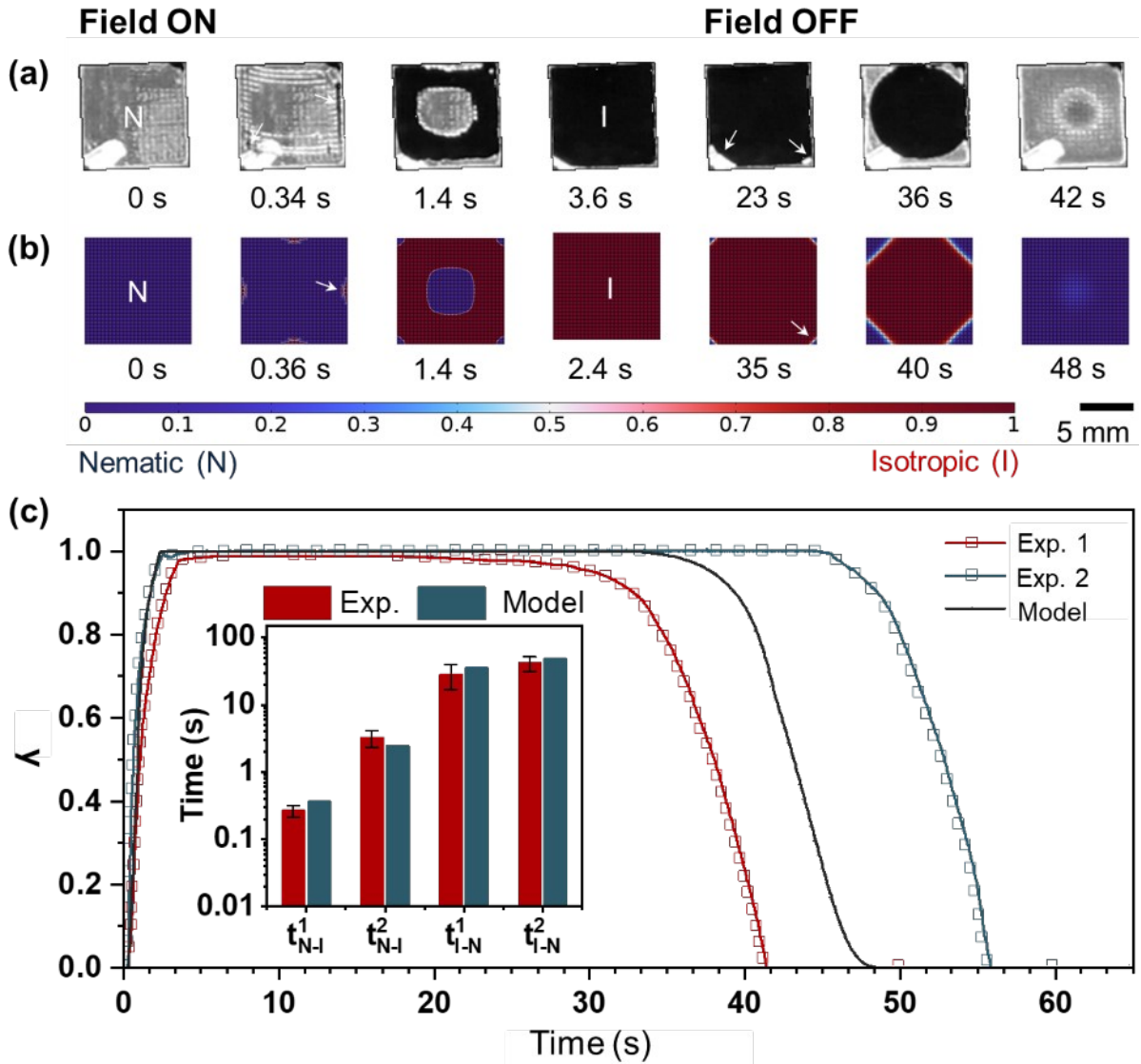

**Figure 3**: (a) Experimental and (b) computational visualization of the phase transition of the sensor chip under a high-frequency magnetic field (30 mT, 125 kHz) triggering optical switching. (c) Reversible N-I transient behavior of the chip, quantified by the ratio of I-phase area to the total chip area ($Λ = A_I/A_T$), obtained from experiments (two representative examples) and computational modeling. The inset shows the corresponding response time parameters of the sensor chip.

A computational model was developed to simulate the phase transition dynamics of the hybrid LC matrix under alternating magnetic fields. Fig. 3b shows the predicted temporal response of the system, which aligns closely with the experimental data. Fig. 3 provides both qualitative and quantitative comparisons between numerical predictions and experimental observations, demonstrating that the simulations generally fall within the standard deviation of the measured data. Any minor discrepancies may be attributed to experimental factors, including variations in activation duration, environmental conditions, and the LC volume fraction within the cell.

The transient response of the LC chip in an alternating field is qualified using the area ratio of I-phase to total projected area $\left(\lambda = \dfrac{A_I}{A_T}\right)$. Performance metrics of the LC chip are extracted from the data as follows:

- Time to initiate the nematic-to-isotropic transition ($t^1_{N\text{-}I}$): Approximated as the time point when $\lambda = 0.01$ during the N-I phase transition.
- Time to complete the nematic-to-isotropic transition ($t^2_{N\text{-}I}$): Determined as the time point when $\lambda \geq 0.99$.
- Time to initiate the isotropic-to-nematic transition ($t^1_{I\text{-}N}$): Defined as the time point when $\lambda \leq 0.99$ during the I-N phase transition.
- Time to complete the isotropic-to-nematic transition ($t^2_{I\text{-}N}$): Approximated as the time point when $\lambda \leq 0.01$ during the I-N phase transition.

An experimental investigation was conducted to evaluate the response of an LC chip equipped with a 50-mesh Ni gauze to the EMF with varying field amplitudes at a frequency of 125 kHz. The results are presented in Figs. 4, 5a and 5b. Time parameters associated with the N-I transition decreased as the amplitude of the applied field increased, whereas the parameters for the I-N transition increased, albeit with a shallower slope. Specifically, $t^1_{N\text{-}I}$ and $t^2_{N\text{-}I}$, representing the N-I transition, increased gradually with increasing field amplitude. This behavior may be attributed to the fact that the I-N transition is governed by the amount of heat generated in the LC cell during field application and the subsequent heat transfer/dissipation to the surroundings. Numerical predictions of these transition times were compared to the experimental data in Figs. 5a and b. The computational model accurately predicts $t^2_{I\text{-}N}$ within experimental error and $t^1_{I\text{-}N}$ for fields up to 20 mT. For field amplitudes above

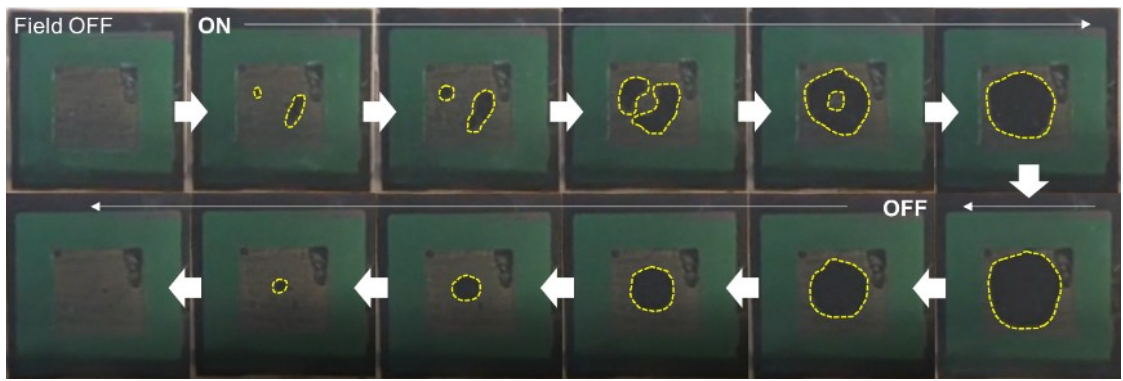

**Figure 4:** Time-lapse images showing the high-frequency magnetic field sensing of the chip depicted in Fig. 2. Upon application of a field (125 kHz, 90 mT), optically dark domains (highlighted with yellow dotted lines) grow near the chip center and rapidly spread/merge to cover most of the chip. When the field is removed, the whole dark domain gradually shrinks and eventually disappears. The full cycle occurs approximately 10 s at room temperature. This experiment was conducted under ambient daylight conditions. [Adapted from the authors' relevant patent (Mohizin & Sung, 2026)]

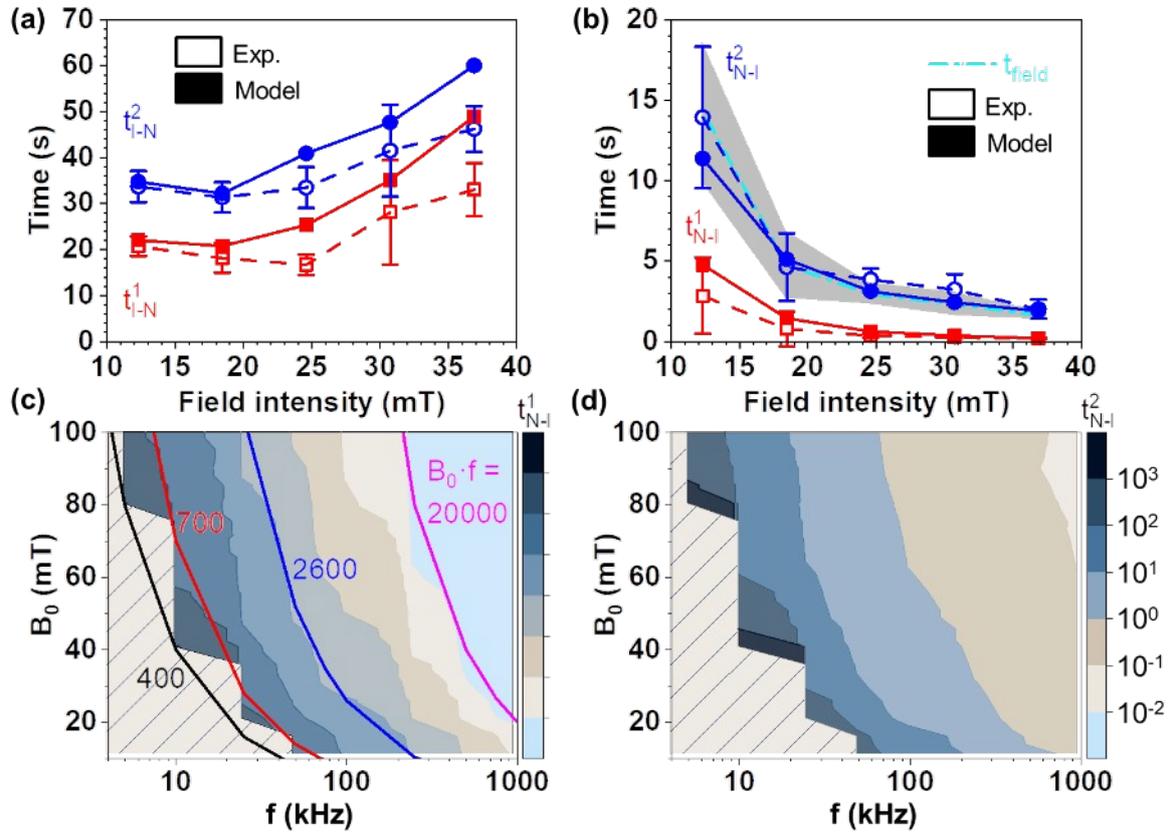

**Figure 5**: Effects of field intensity on the chip's response time parameters for (a) I-N and (b) N-I phase transitions. Computationally derived state diagrams show the response as a function of the field intensity and frequency for (c) $t^1_{I-N}$, and (d) $t^2_{I-N}$.

20 mT, $t^1_{I-N}$ is slightly underestimated by 10-15 s. The model was subsequently used to generate a sensitivity map of the LC chip across varying field intensities and frequencies, with the corresponding data presented in Figs. 5c and 5d.

The effect of mesh geometry on LC chip performance was also examined experimentally, and the results are shown in Fig. 5 and Fig. S2. The volume fraction of the LC cell, defined as the ratio of the Ni mesh volume to the total volume of mesh plus 5CB, was used to compare chip performance under a 30-mT and 125-kHz field. The data, summarized in Fig. 6, indicate that the sensitivity of the LC chip increases with the Ni volume fraction.

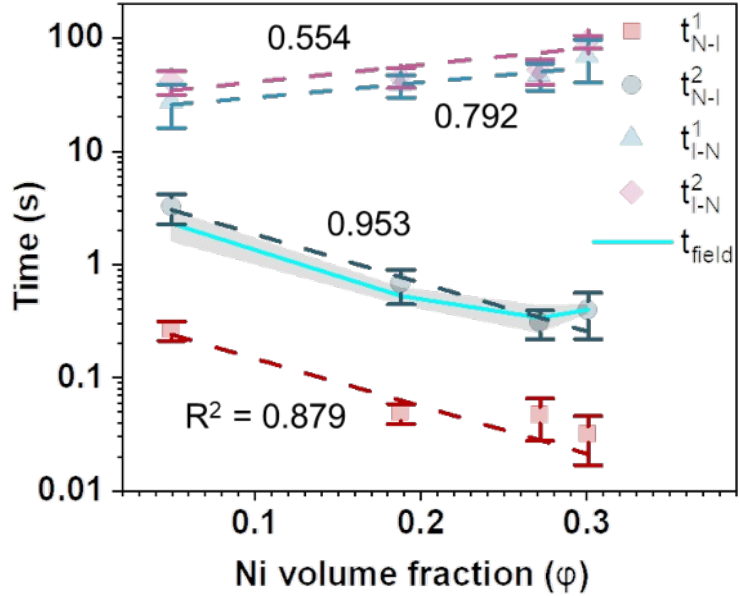

**Figure 6**: Dependence of the chip's response time parameters on the Ni volume fraction. Dashed lines indicate linear fits for $t^1_{N-I}$, $t^2_{N-I}$, $t^1_{I-N}$, and $t^2_{I-N}$, represented by the following relationships: $\log(t^1_{N-I}) = -0.41 - 4.205 f$; $\log(t^2_{N-I}) = 0.696 - 4.286 f$; $\log(t^1_{I-N}) = 1.344 + 1.322 f$; and $\log(t^2_{I-N}) = 1.464 + 1.503 f$, respectively.

## Discussion

### Device characteristics

The optical response of the chip is robust and highly sensitive, and it can be observed with the naked eye under normal lighting conditions (such as ambient daylight or low-intensity illumination from light emitting diodes) without or with minimal reliance on an external power supply (Fig. 4). The sensitivity to field strength can be tuned by adjusting the ferromagnetic mesh size and/or wire diameter, which modulates the heat generated via magnetic induction. Moreover, the chip's small dimensions (down to millimeter length or width) and low mechanical stiffness (Young's modulus on the order of tens of kPa) provide broad adaptability, making the system highly suitable for integration into wearable or mobile platforms.

### Inductive heating regime

The electromagnetic force (*emf*) generated by the time-varying applied magnetic field produces heating power that is proportional to the square of the *emf* divided by the Ohmic resistance of the mesh. The *emf* itself increases linearly with both the magnetic field strength $B$ and the frequency $f$. Therefore, at low frequencies, the heating power is proportional to $(Bf)^2$. At higher frequencies, however, the induced current is confined to the outer layer of the conductor, characterized by the skin depth:

$$\delta = \sqrt{\frac{2}{2\pi f \, \sigma \mu_r \mu_0}}$$

where $\sigma$ is the conductivity of the Ni wire (S/m) and $\mu_r$ is its relative permeability. As the frequency increases, the skin depth decreases, which increases the mesh resistance and reduces the heating power.

Using literature values for the material parameters (Table 1), we estimate a skin depth of 30 μm at our experimental frequency of 125 kHz, which is on the same order as the wire diameter of 50 μm. Consequently, our experimental setup does not clearly fall into either the low- or high-frequency regime. Moreover, the drive frequency of our field generator cannot be easily adjusted. COMSOL simulations (Figs. 5c and d) indicates that the onset of the N-I transition follows a line of constant $B_0 f$ (shown as colored solid lines in Fig. 5c). Assuming that the transition onset is proportional to the power generated in the mesh, this suggests that, at least in theory, our system operates in the low-frequency regime. Increasing the wire diameter could therefore enhance the heating power, but would reduce the mechanical flexibility of the device.

*Effects of ferromagnetic mesh materials*

A ferromagnetic mesh was selected due to its potential for magnetic field enhancement. Because magnetic heating power is proportional to the square of the field intensity, the magnetothermal effect was expected to play a significant role in the sensor design. Alternative materials with higher relative permeability could also be considered. It should be noted, however that both the material's relative permeability and the demagnetizing fields within the wire mesh influence the overall response (Chen et al., 2025). These demagnetization fields are difficult to predict due to the complex mesh geometry and the generation of secondary magnetic fields by eddy currents.

In our experiments, the global orientation of the rod-like 5CB molecules within the LC cell was not pre-aligned, consistent with conditions in our previous work (Sung et al., 2020). It is conceivable that the ferromagnetic wire could be coated with a material such as hexadecyltrimethylammonium bromide (CTAB). By coating the wire surface, the orientation of 5CB rods in the LC matrix could be better controlled through anchoring on the wires. The effects of controlled LC alignment on sensor performance remain an area for further investigation.

*Sensor response time*

The response time for the LCs to undergo a phase transition upon exceeding the N-I transition temperature ($T_{N-I}$) is influenced by several key factors, including the type and density of the thermosensitive LC and the grid structure of the embedded ferromagnetic mesh. Our results show that the response time decreases with increasing mesh size, likely due to the higher volume fraction of ferromagnetic material, which leads to increased heating power. Additionally, reducing the mesh size accelerates heat transfer from the ferromagnetic wires to the LC matrix, further decreasing the response time.

If the response time exceeds 20 seconds, it indicates that the system's efficiency in responding to high-frequency EMFs is compromised. In such cases, modifications to the sensor design—such as adjusting the LC composition, tuning the mesh structure, or altering the LC density—can optimize the response time. By refining these parameters, it is possible to achieve a sensor response with the desired performance range, thereby enhancing the overall efficiency and applicability of the system under varying radio-frequency EMF conditions.

*Phase transition temperature of the LC matrix*

For the LC material used in this work, any thermotropic LC compounds exhibiting reversible N-I transition can be employed. Ideally, the N-I transition should fall within a temperature range of 30-50°C, just above room temperature, yet not so high as to risk damage to the device's polymer components. In general, the choice of the LC material should match the intended operating temperature range of the sensor. For example, a lower $T_{N-I}$ would enable applications at lower temperatures, such as in aerospace technology, whereas a higher $T_{N-I}$ would allow sensor operation at elevated temperatures.

For applications at ambient temperatures in the temperate zones, the thermosensitive LC material may comprise one or more LC compounds, such as N-(p-methoxybenzylidene)-p'-n-butylaniline (MBBA; $T_{N-I}$ = 47°C) or p-pentylphenyl-trans-p'-pentylcyclohexylcarboxylate ($T_{N-I}$ = 47°C), in addition to 5CB used in the present study.

**Conclusion**

We developed a novel sensor chip for detecting high-frequency alternating magnetic fields. The sensor consists of a matrix of thermosensitive LC, which undergos a reversible nematic-isotropic phase transition, and a mesh of ferromagnetic wires in thermal contact with the LC matrix, forming a thin sandwich structure confined between two crossed polarizing films. The device operates without any electrical connections, functioning as a field-responsive LC microdisplay.

Experiments conducted at 125 kHz with field amplitudes up to 30 mT demonstrated a response time as fast as 30 ms upon induction heating, with the LC phase transition propagating from the exterior toward interior. Both the response time and the phase transition pattern are in good agreement with finite element simulations within the experimental error. The heating-induced optical response time decreases with increasing mesh volume fraction, whereas the recovery time—on the order of tens of seconds—is longer and could not be precisely captured by the simulations. Notably, the recovery time increases with increasing mesh volume fraction, in contrast to the heating response.

The combination of extreme simplicity, passive operation, and rapid responsiveness makes this sensor a promising candidate for applications in ultra-low-power wearable technologies, mobile platforms, and electronic warfare systems.

**Acknowledgements**

This work was supported by the KIST-Europe Basic Research Program (Project No. 12320 & 12402) and the National Research Council of Science & Technology (Project No. Global-22-015) of South Korea. We thank Dr. Matthias Altmeyer (KIST Europe) for his technical assistance in operating and managing the high-frequency field generator.

# Electronics-free, ultra-low-power, wearable sensor chip for high-frequency electromagnetic field detection

Abdul Mohizin[1†], Leon Abelmann[2†], Baeckkyoung Sung[1,3*]

1. Biosensor Group, KIST Europe Forschungsgesellschaft mbH, 66123 Saarbrücken, Germany
2. Department of Microelectronics, Delft University of Technology, 2628 CD Delft, The Netherlands
3. Division of Energy & Environment Technology, University of Science & Technology (UST), 34113 Daejeon, Republic of Korea

† These authors contributed equally to this work.
*Correspondence: sung@kist-europe.de

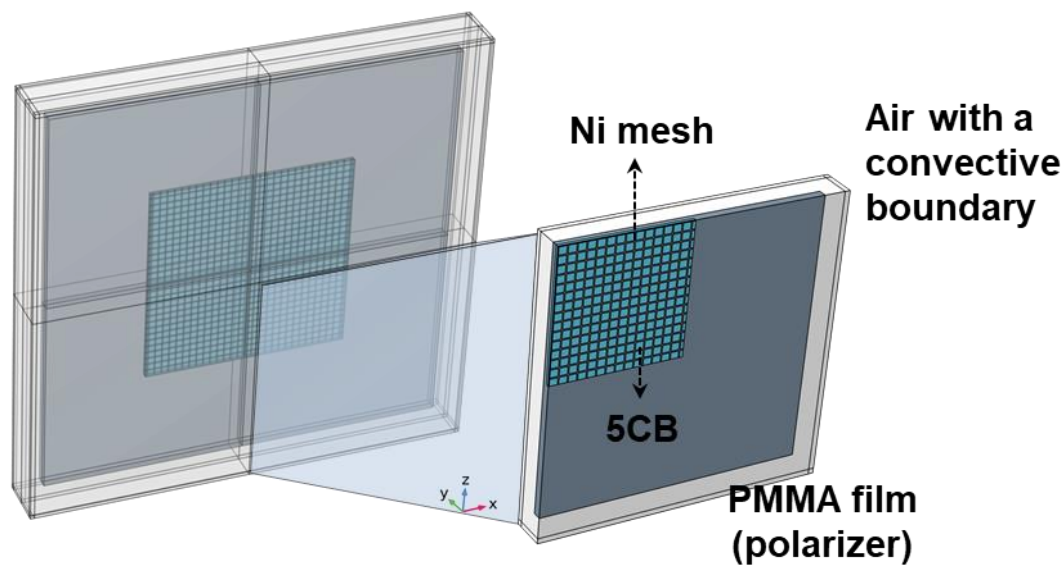

**Figure S1**: 3D computational domain of the sensor chip used in the COMSOL modeling. The entire geometry was divided into 8 identical segments, where only one segment was used for the simulation study with symmetric boundary conditions. The phase transition of thermotropic liquid crystal (5CB) is driven by magnetothermal effect induced by high-frequency magnetic fields.

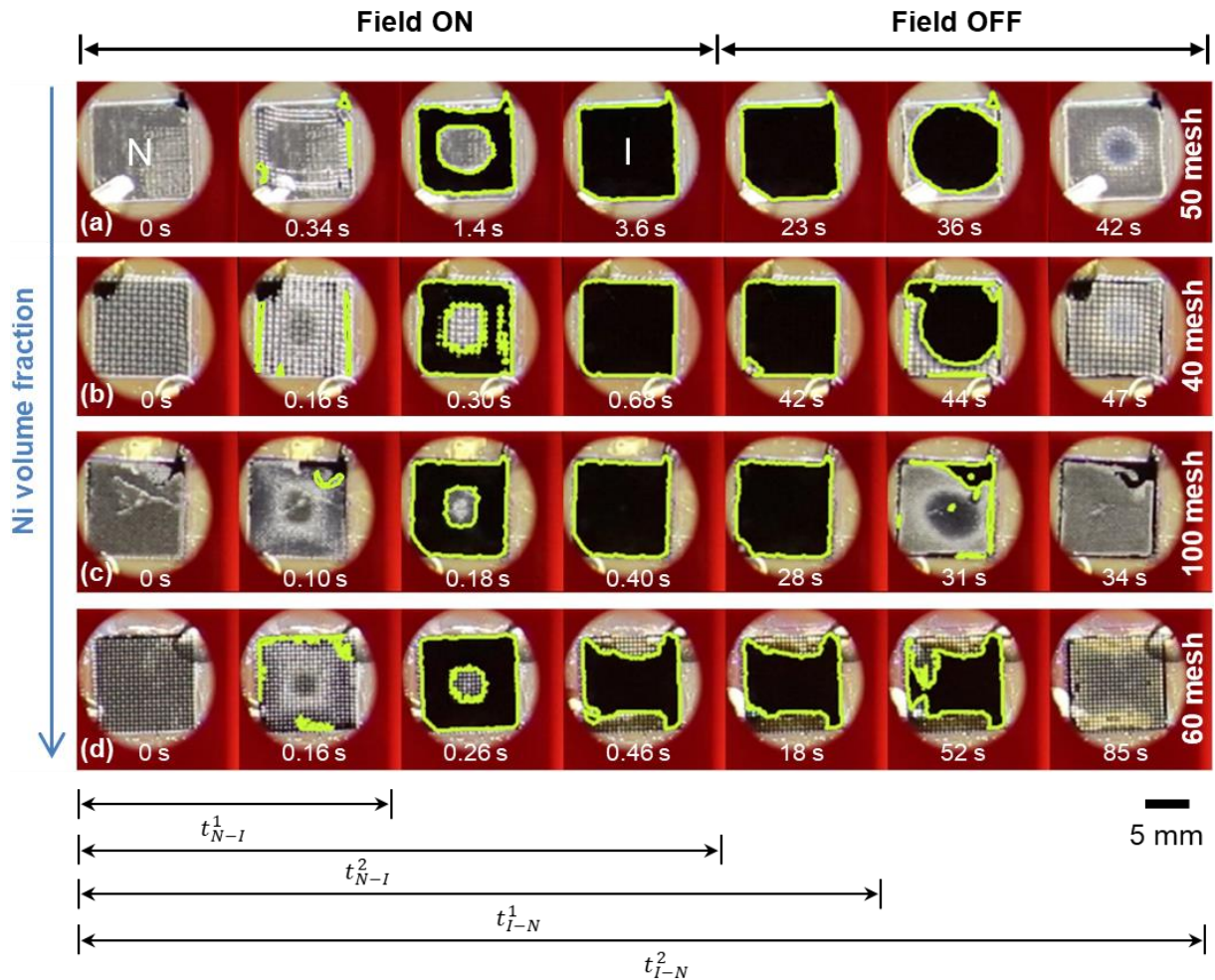

**Figure S2**: Representative images showing the optical responses of sensor chips (caused by a reversible isotropic-nematic phase transition of thermoresponsive 5CB layer) in different conditions of Ni volume fraction and mesh size. From the upper to lower rows, the Ni volume fraction increases from (a) 0.050 (50 mesh), (b) 0.188 (40 mesh), (c) 0.272 (100 mesh), to (d) 0.301 (60 mesh). [N: nematic phase, I: isotropic phase]

**Movie S1**: A representative movie demonstrating the operation of a sensor chip under a pulse of high-frequency magnetic field. The field application (the onset represented by an aluminum piece' blow-away) continued for about 10 s. The sensor chip was located on the top of support (in the middle of the Helmholtz coil) and its 90°-reflected images were continuously obtained with time evolution (Fig. 2).